\documentclass[aps,showpacs,showkeys,preprint,amsmath,amssymb]{revtex4}

\usepackage{natbib}  % Phil. Mag. citation style
\bibpunct{(}{)}{,}{a}{}{,} % Phil. Mag. citation style

\usepackage{graphicx}% Include figure files
\usepackage{dcolumn}% Align table columns on decimal point
\usepackage{bm}% bold math
 
\usepackage{epsfig}

\begin{document}
%\draft

\title{Length-scale-dependent phase transition \\ 
in $\rm{Bi_{2}Sr_{2}CaCu_{2}O_{8}}$ single crystals}

\author{K. Vad} 
\email{vad@atomki.hu}
\affiliation{Institute of Nuclear Research of the Hungarian Academy 
of Sciences, H-4001 Debrecen, P.O.Box 51 Hungary
}%

\author{S. M\'esz\'aros}
\affiliation{Institute of Nuclear Research of the Hungarian Academy 
of Sciences, H-4001 Debrecen, P.O.Box 51 Hungary
}%

\author{I. N\'andori} 
\affiliation{Institute of Nuclear Research of the Hungarian Academy 
of Sciences, H-4001 Debrecen, P.O.Box 51 Hungary
}%

\author{B. Sas}
\affiliation{ Research Institute of Solid State Physics, P.O.Box 49, 
H--1525 Budapest, Hungary
}%

\date{\today}

\begin{abstract}
Electrical transport measurements using a multiterminal configuration are
presented, which prove that in
${\rm{Bi_{2}Sr_{2}CaCu_{2}O_{8}}}$ single crystals near the
transition temperature in zero external magnetic field the secondary voltage
is induced by thermally activated vortex loop unbinding. The phase
transition between the bound and unbound states of the vortex loops was
found to be below the temperature where the phase coherence of the
superconducting order parameter extends over the whole volume of the
sample. We show experimentally that 3D/2D phase transition in vortex
dimensionality is a length-scale-dependent layer decoupling process and
takes place simultaneously with the 3D/2D phase transition in
superconductivity at the same temperature.
\end{abstract}

\maketitle

\section{Introduction}
Of high-$T_c$ superconducting compounds
$\rm{Bi_{2}Sr_{2}CaCu_{2}O_{8}}$ is preferred because its discrete
superconducting layers play an important role in current conduction
properties even in the transition temperature range. The multicontact
configuration is one of the most promising arrangements of electrical
transport measurements in the study of superconducting phase transition
and vortex dimensionality. In this configuration four electrical contacts are
attached to one side of a single crystal and two or four contacts to the
opposite side. If the current injected into one side of a crystal is high
enough, a voltage drop on both sides can be measured. The side where the
current is injected is called primary, the opposite side is secondary. Due to
this contact arrangement and the anisotropic conductivity, in
$\rm{Bi_{2}Sr_{2}CaCu_{2}O_{8}}$ a non-uniform current distribution
develops, and the current injected into the primary side is confined to a
very thin surface layer, which is thinner than the thickness of the crystal. If
the pancakes in adjacent layers belonging to the same vortex line are
strongly coupled to each other, the primary current induces vortex motion
both in the primary and in the secondary surface. Moreover, if the coupling
strength between the pancakes is high enough to produce three-dimensional
(3D) type vortex lines across the sample, primary and secondary voltages
can be equal. This configuration is called dc flux transformer configuration.

The first measurements on $\rm{Bi_{2}Sr_{2}CaCu_{2}O_{8}}$ single
crystals using the dc flux transformer configuration were performed by  
\citet{safar} and  \citet{busch}.
They observed the dimensionality of the vortex system over a wide range
of the phase diagram. \citet{wan1,wan2} and C.D. \citet{keener} also used 
the multicontact dc flux
transformer configuration to study the secondary voltage and found that in
magnetic fields near the transition temperature the interlayer vortex coupling
was responsible for the secondary voltage. However, the origin of the
secondary voltage in zero external magnetic field still remained problematic.
\citet{pierson1} suggested that it originated from thermally
activated vortex loop unbinding. A vortex loop is a correlated
vortex-antivortex line pair. Using a real-space renormalization group
analysis, the author identified three characteristic critical currents and
calculated their temperature dependence. He found that the temperature
dependence of the secondary voltage is a horizontal slice of the
current-temperature phase diagram.

In this paper we present electrical transport measurements using a
multiterminal configuration that prove the 3D/2D phase transition in vortex
dimensionality near the Ginzburg-Landau transition temperature and show
that this phase transition is a length-scale-dependent layer decoupling
process. We show that the temperature dependence of the secondary
voltage in zero magnetic field has a double peak structure.

\section{Experimental arrangement}
Single-crystalline $\rm{Bi_{2}Sr_{2}CaCu_{2}O_{8}}$ compounds
were prepared by the melt cooling technique, described in
Reference \citep{keszei,keszei2}. Optically smooth rectangular crystals were
carefully cleaved from these compounds, and heated in flowing oxygen for
15 minutes at 900 K in order to stabilize the oxygen content. Chemical
homogeneity of the samples was checked by microbeam PIXE and
(O$^{16}, \alpha $) resonant elastic scattering \citep{simon,simon2} with a spatial
resolution of 5 $\mu $m. No chemical inhomogeneities were identified. The
surface smoothness was measured by atomic force microscopy. The
surfaces were found to be flat with a typical roughness of 10 nm. Before
preparing the electrical contacts, the sample quality was checked by
magnetization measurements using a SQUID or a vibrating sample
magnetometer, and by AC susceptibility measurements. Electrical contacts
were made by bonding 25 $\mu $m gold wires with Dupont 6838 silver
epoxy fired for five minutes at 900 K. The contact resistance was a few
ohms. The geometrical position of electrical contacts was precisely
measured by optical microscope.

\begin{figure}
\epsfig{file=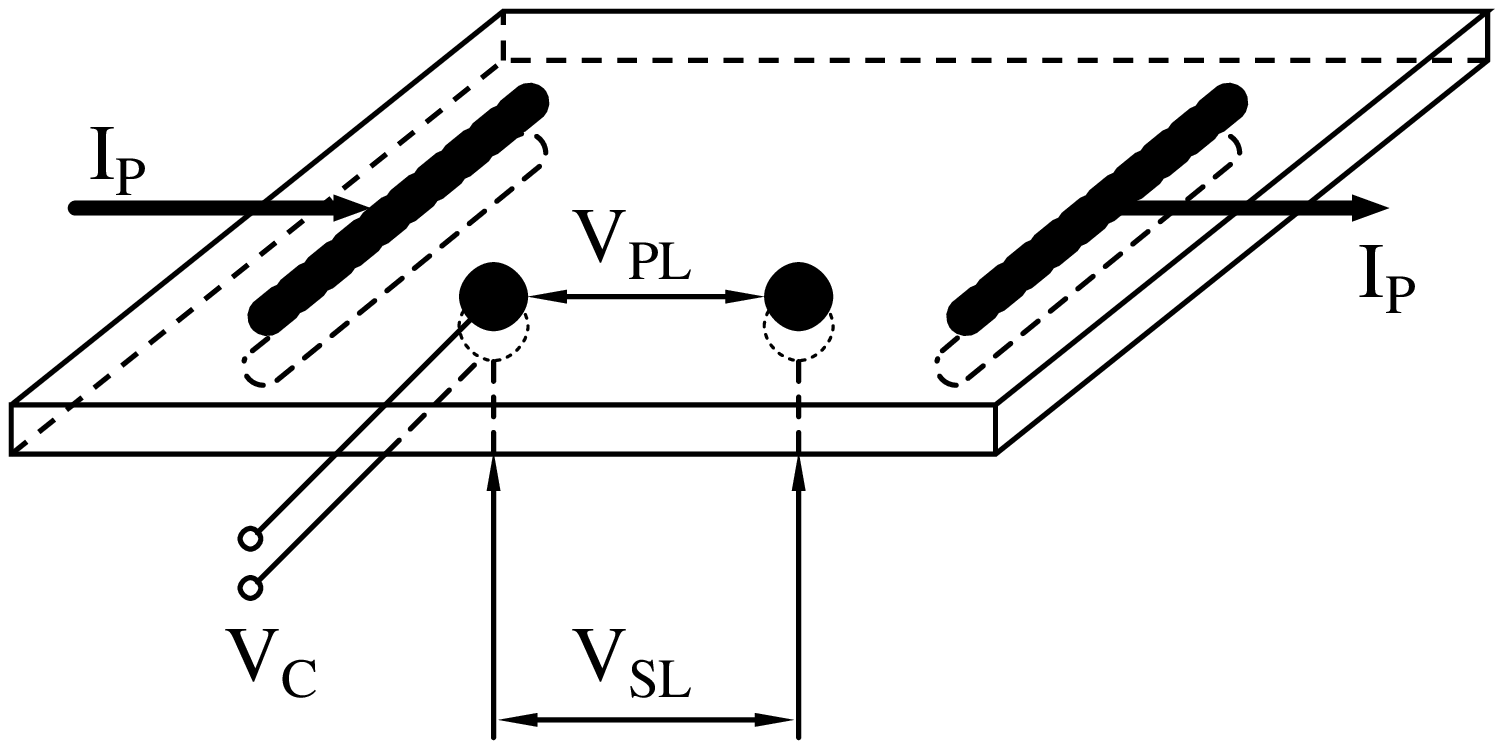, width=6cm}
\caption{Electrode configuration used for secondary voltage
measurements. $V_{PL}$ and $V_{SL}$ denote the primary and
secondary longitudinal voltages, $V_{C}$ and $I_P$ denote the $c$-axis
direction voltage and the primary current, respectively.}
\label{fig1}
\end{figure}

Two current and potential electrodes were attached to both faces of
the crystals. The scheme of the electrode configuration is shown in
Fig.\ \ref{fig1}. The current was injected into one face of the single crystal
through the current contacts (this is the primary current $I_P$), while
primary and secondary longitudinal voltages, and the voltage between the
two faces were recorded simultaneously.

We have performed measurements on a few crystals fabricated from the
same batch. The mean-field Ginzburg-Landau (GL) transition temperature
$T_{c0}$ of the samples was 88 K. The sample dimensions were
about $1\times 1.5$ mm$^2$, the thickness was between 8 and 3 $\mu $m.
As we pointed out in our previous papers \citep{sas,sas2}, most of the Joule heat
due to the transport current was generated in the current contacts. In order
to reduce the heat dissipation, and also to eliminate the thermoelectric
force, we used current pulses with a duration time of 1 ms and repetition
time of 100 ms. This arrangement and the fact that the sample temperature
was regulated by a temperature controlled He gas stream instead of
exchange gas, made it possible to avoid heating during the pulse up to the
amplitude of 10 mA. Using the analysis of 
\citet{busch}, we could define the temperature dependence of the $ab$
plane and $c$-axis resistivities from the voltages $V_{PL}$ and
$V_{SL}$. For the anisotropy ratio $\gamma$ we received
$\gamma =\sqrt(\rho _{c}/\rho _{ab}) \approx 500$ with
$\rho _{ab} \approx 100$ $\mu \Omega$cm at 90 K.

\section{Measurements}
In our experiments the following voltages, shown in Fig.\ \ref{fig1}, were
simultaneously recorded: (i) longitudinal voltages measured parallel to the
current direction on the surface of the crystal where the current was
injected and on the opposite surface, i.e., primary and secondary
longitudinal voltages ($V_{PL}$ and $V_{SL}$); (ii) the $c$-axis
direction voltage measured between the two surfaces of the crystal
($V_{C}$).

The temperature dependence of the primary and secondary longitudinal
voltages of a ${\rm{Bi_{2}Sr_{2} CaCu_{2}O_{8}}}$ single crystal
measured in zero magnetic field is shown in Fig.\ \ref{fig2}.
In zero applied magnetic field the secondary voltage is determined by
interlayer vortex coupling. In order to gain information about this coupling
strength, we measured the voltage $V_C$ between the two surfaces of the
crystal while applying a primary current $I_P$. Depending on the current
distribution in the crystal, some part of this current flows in the $c$-axis
direction and $V_C$ depends on this current. The temperature
dependence of $V_C$ is of metallic ($d\rho/dT>0$) type, except in a small
temperature range near the mean-field Ginzburg-Landau transition
temperature. In Fig.\ \ref{fig3}(a) this range is between 85 and 86 K. This
temperature dependence is also reflected in the current-voltage
characteristics (Fig.\ \ref {fig3}(b)). Far above the GL transition
temperature, at 256 K the current-voltage characteristic is linear. Near the
GL transition temperature, at 87.3 K there is a slight curvature in it. On
further cooling the sample from 87.3 K to 86 K the curvature increases. In
the temperature range between 86 K and 85 K the $I_{P}-V_C$ curves are
concave in shape (for clarity in Fig.\ \ref{fig3} we present only the curve
measured at 85.8 K). At temperatures lower than 85 K both the
$V_{C}-I_{P}$ and $V_{C}-T$ curves show the same metallic behavior,
as above 86 K.

\section{Results and discussion}
In 2D superconducting layers the phase fluctuation of the order parameter
generates vortex-antivortex pairs as topological excitations \citep{tinkham}.
The phase transition in an isolated 2D superconducting layer, where the
vortex-antivortex pairs are bound below the phase transition temperature
and are unbound above it, is described by the Kosterlitz-Thouless theory
\citep{kosterlitz,halperin}. In thin films, if the film thickness is less than the
superconducting coherence length, this Kosterlitz-Thouless phase
transition ($T_{KT}$) can be observed.

In high-temperature superconductors, the high transition temperature, short
coherence length and layered structure make the phase fluctuation of the
order parameter dominant over the other fluctuations in the transition
temperature range, but the coupling between superconducting layers
modifies the 2D Kosterlitz-Thouless picture. The interaction between
superconducting layers leads to vortex-antivortex interaction different from
the 2D case. In $\rm{Bi_{2}Sr_{2}CaCu_{2}O_{8}}$ single crystals the
CuO$_2$ planes serve as 2D superconductor layers, the vortices are the
pancake vortices. Due to the layered structure a high anisotropy exists in
conductivity and the vortex matter has a very rich phase diagram with
numerous phase transitions. Among others the 3D phase appears, where
the coupling between neighboring layers arranges the thermally excited
pancake vortices into 3D flux lines \citep{blatter}. These 3D flux lines can
form vortex loops as correlated vortex-antivortex line pairs. They are called
thermally activated vortex loops, because they are the result of a combined
effect of thermally activated vortex excitation and interlayer vortex
coupling.

The 3D character modifies the structure of the phase transition between the
bound and unbound states: a narrow 3D window appears around the phase
transition temperature ($T_c$) and a nonzero critical current appears. The
3D temperature region is theoretically predicted and it has been shown that
the size of this 3D region is much smaller above $T_c$ than
below \citep{pierson2}. Above the phase transition temperature the behavior
of vortices becomes 2D due to decoupling of the superconducting layers
(3D/2D phase transition). This layer decoupling is a length-scale-dependent
process: the layers become decoupled at length scales larger than an
interlayer screening length, while for lengths below this scale they remain
coupled.

In an isolated 2D superconducting layer the vortex pair energy is the
sum of the creation energy of the two vortices and the intralayer logarithmic
coupling energy between vortices. In layered systems like
$\rm{Bi_{2}Sr_{2}CaCu_{2}O_{8}}$ this vortex pair energy is modified
by the interlayer Josephson coupling, which not only strengthens the
intralayer interaction, but causes an interaction between vortices located in
neighboring layers. The intralayer vortex pair energy can be written in the
following simple form:
$E(r)=2E_{c}+K_{\parallel} ln(r/\xi_{0})+K_{\perp} r/\xi_{0}$,
where $r$ is the distance between the vortex and antivortex, $E_{c}$ is the
creation energy of a vortex, $\xi_{0}$ is the zero-temperature correlation
length, $K_{\parallel}$ is the intralayer vortex-vortex coupling constant and
$K_{\perp}$ is the interlayer Josephson coupling constant. While the 2D
logarithmic interaction (second term in this equation) dominates at short
distances, the Josephson-coupling mediated 3D linear interaction (third
term) dominates at large distances. The intralayer vortex length scale
$R_{\lambda}(T)$ is the characteristic length which divides this
logarithmic and linear regimes \citep{pierson3}.
$R_{\lambda}(T)=\xi_{0}/\sqrt \lambda$, where $\lambda$ is the ratio of
the interlayer Josephson coupling to the intralayer coupling,
$\lambda=K_{\perp}/K_{\parallel}$. $\lambda$ depends on the size $r$ of
the vortex pairs, and above $T_c$ it has a maximum. The size $r$ which
belongs to this maximum $\lambda$ is the other characteristic length, the
interlayer screening length ${\it l}\,_{3D/2D}(T)$. If the separation
between two vortices located in neighboring layers is larger than
${\it l}\,_{3D/2D}$, the Josephson-coupling mediated linear interaction is
screened out and the layers are decoupled. Around $T_c$ the dimensional
behavior of the system is determined by these two characteristic lengths.
The layers are coupled and the behavior of vortices is 3D if the separation
between vortices is greater than the intralayer vortex length scale
$r>R_{\lambda}(T)$ and less than the interlayer screening length
$r<{\it l}\,_{3D/2D}(T)$. The temperature dependence of
$R_{\lambda}(T)$ and ${\it l}\,_{3D/2D}(T)$ shows \citep{pierson4} that
the intralayer vortex length scale $R_{\lambda}(T)$ is constant for
$T \ll T_c$, but it increases as the temperature approaches $T_c$ from
below. The interlayer screening length ${\it l}\,_{3D/2D}(T)$ decreases
continuously as the temperature increases.

While in the 3D regime below $T_{c}$ the electrical transport behavior is
dominated by vortex loops, above $T_{c}$ it is dominated by vortex lines
and pancake vortices. The multiterminal configuration is a good
arrangement to distinguish between these two regimes, and the temperature
dependence of the secondary voltage can help us to understand the length-
scale-dependent layer decoupling, i.e., the 3D/2D phase transition.

\begin{figure}
\epsfig{file=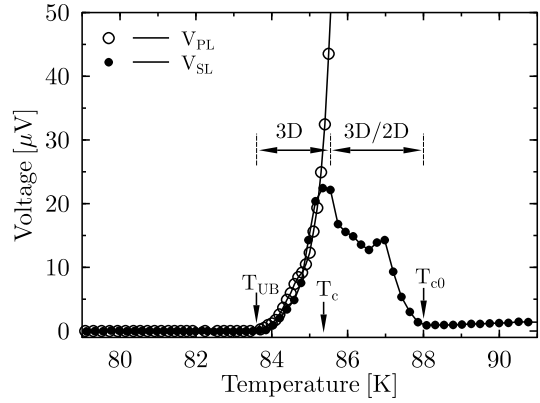, width=7cm}
\caption{Secondary and primary voltages vs. temperature in zero magnetic 
field, $I_{P}$ = 1 mA. $T_{UB}$, $T_c$ and $T_{c0}$ denote the 
unbinding, transition and Ginzburg-Landau transition temperature,  
respectively. The different vortex dimensionalities are also shown.}
\label{fig2}
\end{figure}

\begin{figure}
\epsfig{file=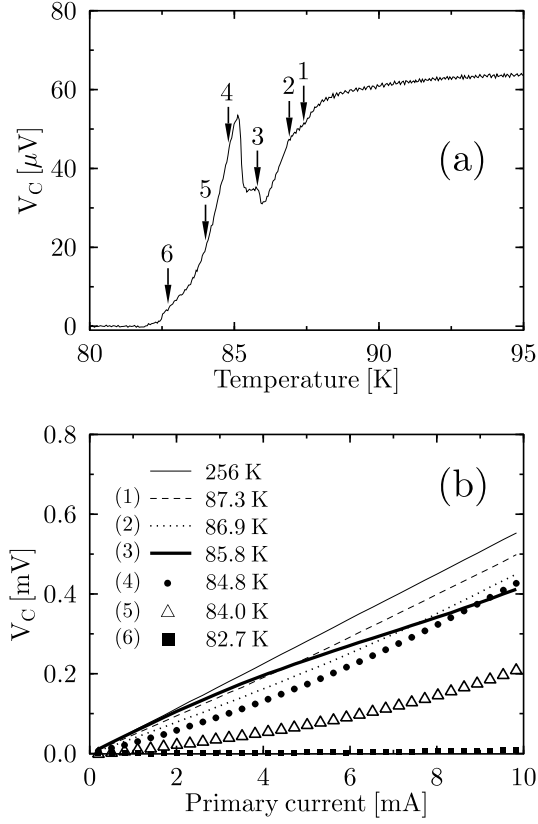, width=7cm}
\caption{Temperature dependence of $V_{C}$ at 1 mA primary current
(a), and current-voltage characteristics measured at different temperatures
(b) in zero magnetic field. The arrows in (a) denote the temperatures where
the current-voltage characteristics were measured.}
\label{fig3}
\end{figure}

The secondary voltage has already been studied both theoretically
\citep{pierson1} and experimentally \citep{wan1}. It can now be accepted that
in zero applied magnetic field it originates from thermally activated vortex
loop unbinding. At low temperatures where $V_{SL}$ and $V_{PL}$ are
zero, the thermally excited 3D flux lines form vortex loops which are
'pinned' to the crystal. With increasing temperature, the transport current
splits these vortex loops into free vortex-antivortex line pairs. The
temperature where this splitting starts is the unbinding temperature
($T_{UB}$). This is the lowest temperature where both $V_{SL}$ and
$V_{PL}$ are observable. Above the unbinding temperature the free
vortices move in the sample like 3D vortex lines due to the Lorentz force,
producing the same voltage drop on the primary and secondary side of the
crystal, $V_{PL}=V_{SL}$. This 3D character of the vortex lines remains
up to a temperature where the secondary voltage has a local maximum. In
Fig.\ \ref{fig2} the 3D temperature range is around 85 K where the primary
and secondary longitudinal voltages $V_{PL}$ and $V_{SL}$ coincide.
This is the same 3D temperature range which was predicted theoretically by
renormalization group analysis \citep{pierson2}. With increasing temperature
the 3D character of flux motion disappears and consequently $V_{SL}$
becomes lower than $V_{PL}$, but another local maximum of $V_{SL}$
can be found as the temperature approaches $T_{c0}$. The temperature
dependence of the secondary voltage has two peaks with a higher and a
lower amplitude.

\begin{figure}
\epsfig{file=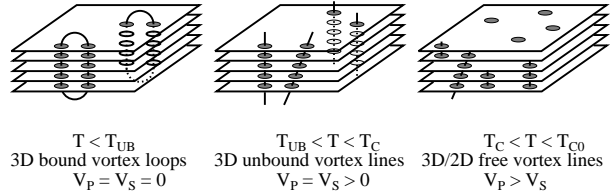, width=8cm}
\caption{Schematic illustration of different vortex types.}
\label{fig4}
\end{figure}

This double peak structure of $V_{SL}$ can be explained by the
motion of different types of vortex lines. In zero applied magnetic field free
vortex lines can be produced in two ways. First, they can be the result
of vortex-antivortex depairing of thermally activated vortex loops due
to the Lorentz force of the transport current. In this case the number of
free vortex lines depends on the transport current density and a non-Ohmic
behavior characterizes the system. Secondly, free vortex lines can be
spontaneously created by thermal activation, mainly above $T_{c}$. The
number of free vortex lines increases with increasing temperature and the
system is characterized by an Ohmic behaviour. In Fig.\ \ref{fig4} we
illustrate the different types of vortices and vortex dimensionalities
schematically.

The effect of current on vortex-antivortex depairing is twofold. On the one
hand the current reduces the creation energy whereby increases the density
of vortex pairs. On the other hand the current exerts a force (the Lorentz
force) on vortex loops and can blow them out. The number of blowouts
depends on the size $r$ of the vortex pairs. If $r$ is higher than the
intralayer vortex length scale $R_{\lambda}(T)$, the Josephson-coupling
mediated 3D linear interaction is energetically favored over 2D intralayer
logarithmic interaction and the energy of a vortex loop is smaller than the
energy of a pair of independent vortex lines.
Below $T_c $ in $\rm{Bi_{2}Sr_{2}CaCu_{2}O_{8}}$
$R_{\lambda}(T)=\xi_{0}/\sqrt \lambda \approx 1 \mu$m, where
$\xi_{0} \sim 3$nm and $\lambda \sim 10^{-5}$. Consequently, if
$r>1 \mu$m, creation of vortex loops is energetically favored over free
vortex-antivortex line pairs. This happens in the 3D temperature range
where the dominant topological excitation is the vortex loop. With
increasing temperature the number of blowouts and, so, the
secondary voltage increases. However, approaching the transition
temperature the intralayer vortex length scale $R_{\lambda}(T)$ starts to
increase, therefore the number of vortex pairs which can be blown out by
the transport current decreases which results in the decrease of the
secondary voltage. The temperature which belongs to the peak value of the
secondary voltage is the transition temperature $T_c$. Although at this
temperature the behavior of vortices is still 3D type as it was shown
theoretically \citep{pierson4}, the secondary voltage is somewhat lower than
the primary voltage (see Fig.\ \ref{fig2}). This means that the secondary
voltage starts to decrease before layer decoupling. At higher temperatures
the 3D character of flux motion disappears and $V_{SL}$ becomes
significantly lower than $V_{PL}$. Above the 3D temperature range
another local maximum of $V_{SL}$ can be found as the temperature
approaches $T_{c0}$, because the temperature is high enough to
produce free vortex lines by thermal activation. The number of free vortex
lines and, so, the secondary voltage increases with increasing temperature.
The system is characterized by a continuous 3D/2D transition due to
continuous decrease of the interlayer screening length
${\it l}\,_{3D/2D}(T)$ as the temperature approaches the mean-field
Ginzburg-Landau transition temperature $T_{c0}$. Near $T_{c0}$ the
amplitude of the order parameter decreases, just as the number of the free
vortex lines. This effect evokes the decrease of the secondary voltage. The
temperature which belongs to the minimum secondary voltage above the
double peak structure is $T_{c0}$.

Information about the strength of Josephson coupling and the 2D/3D phase
transition in superconductivity can be obtained by study the temperature
dependence of $I_{P}-V_C$. In the temperature range where the
$I_{P}-V_C$ curves are concave (between 85 and 86 K) the
superconducting coherent state exists in the CuO$_2$ bilayers, but the
interlayer coupling is not strong enough to establish the $c$-axis
coherence. In consequence, the current in $c$-axis direction is carried by
single particle tunnelling instead of Cooper pair tunnelling (2D type
superconductivity). 
There is no stabil phase coherence between superconducting bilayers 
and the phase fluctuation is dominant. Consequently, the shape of the 
$V_C -T$ curve in the temperature range where $d\rho/dT < 0$ can slightly 
change at different sample cooling rates. Figure 3 presents the $V_C -T$ 
and  $V_C -I_P$ curves taken at different sample cooling rates. 
With the decrease in temperature, at 85 K, the Josephson coupling, and 
consequently, a phase coherence between superconducting bilayers develops 
and extends over the whole volume of the sample, producing maximum values 
in both $V_{SL}$ and $V_C$ (2D/3D phase transition). At temperatures 
lower than 85 K the Josephson coupling energy increases, the phase 
coherence of the superconducting order parameter extends over the 
whole volume of the sample and develops the 3D type superconductivity. 
The higher peak value of $V_{SL}$ is also at 85 K which is the upper 
end of the temperature range where vortex lines have 3D character. 
Consequently 2D/3D phase transition in superconductivity and in vortex 
dimensionality takes place at the same temperature. This can be valid 
inversely, too. If the vortex dimensionality decreases from 3D to 2D, 
the dimensionality of superconductivity can also decrease. This result 
was experimentally supported by transport current measurements 
\citep{pethes}, where the authors proved that the Bardeen-Stephen model 
for the flux flow resistance $\rho _{f}=\rho _{n}\cdot B/B_{c2}$ 
($\rho _{n}$, $B$ and $B_{c2}$ are the normal state resistivity, 
magnetic field and critical magnetic field) is not valid at high 
current density in $\rm{Bi_{2}Sr_{2}CaCu_{2}O_{8}}$ single crystals. 
Due to intensive flux motion both the phase coherence between 
superconducting bilayers and the interlayer screening length
${\it l}\,_{3D/2D}(T)$ decrease resulting in a 3D/2D phase transition 
in dimensionality of superconductivity.

\section{Conclusions}
In conclusions, we found that temperature dependence of secondary
longitudinal voltage has a double-peak structure and its higher maximum
value is at the temperature where the phase coherence of the order
parameter extends over the whole sample. Secondary
voltage originates from correlated vortex-antivortex line pair unbinding, i.e.,
from vortex loop unbinding due to the Lorentz force of the transport
current. Near $T_{c0}$ free vortex-antivortex line pairs are also generated
by thermally activated vortex excitation. We think that the two types of
vortex-antivortex line pairs are responsible for the double peak structure of
the secondary longitudinal voltage. Lacking of theories describing this
double peak structure in $\rm{Bi_{2}Sr_{2}CaCu_{2}O_{8}}$ single
crystals indicates a need for better description of length-scale dependence
in layered superconductors. In order to improve the theoretical description,
one can perform the renormalization group analysis of Reference
\citep{pierson1} for non-constant current density or one can use different
renormalization group methods, e.g., field theoretical renormalization group
approaches \citep{nandori,nandori2}.

\begin{acknowledgments}
We take great pleasure in acknowledging discussion with P.F. de
Ch\^{a}tel. This work was supported by the Hungarian
Science Foundation (OTKA) under contract no. T037976.
\end{acknowledgments}

\newpage

\end{document}